\documentclass[sigconf,screen]{acmart}

\AtBeginDocument{%
  \providecommand\BibTeX{{%
    \normalfont B\kern-0.5em{\scshape i\kern-0.25em b}\kern-0.8em\TeX}}}

\usepackage{listings}
\usepackage{rotating}
\usepackage{enumitem}
\usepackage{soul}
\usepackage{multirow}
\usepackage[normalem]{ulem}
\usepackage{colortbl}
\usepackage{array}
\usepackage{pifont}
\usepackage{rotating}
\usepackage{tcolorbox}

\copyrightyear{2020}
\acmYear{2020}
\setcopyright{acmcopyright}
\acmConference[ESEC/FSE '20]{Proceedings of the 28th ACM Joint European Software Engineering Conference and Symposium on the Foundations of Software Engineering}{November 8--13, 2020}{Virtual Event, USA}
\acmBooktitle{Proceedings of the 28th ACM Joint European Software Engineering Conference and Symposium on the Foundations of Software Engineering (ESEC/FSE '20), November 8--13, 2020, Virtual Event, USA}
\acmPrice{15.00}
\acmDOI{10.1145/3368089.3409751}
\acmISBN{978-1-4503-7043-1/20/11}
\acmSubmissionID{fse20main-p624-p}

\begin{document}

\title{RulePad: Interactive Authoring of Checkable Design Rules}


\author{Sahar Mehrpour}
\email{smehrpou@gmu.edu}
\affiliation{%
  \institution{George Mason University}
  \city{Fairfax}
  \state{Virginia}
  \country{USA}
}

\author{Thomas D. LaToza}
\email{tlatoza@gmu.edu}
\affiliation{%
  \institution{George Mason University}
  \city{Fairfax}
  \state{Virginia}
  \country{USA}
}

\author{Hamed Sarvari}
\email{hsarvari@gmu.edu}
\affiliation{%
  \institution{George Mason University}
  \city{Fairfax}
  \state{Virginia}
  \country{USA}
}


\begin{abstract}
Good documentation offers the promise of enabling developers to easily understand design decisions. Unfortunately, in practice, design documents are often rarely updated, becoming inaccurate, incomplete, and untrustworthy. A better solution is to enable developers to write down design rules which are checked against code for consistency. But existing rule checkers require learning specialized query languages or program analysis frameworks, creating a barrier to writing project-specific rules. 
We introduce two new techniques for authoring design rules: snippet-based authoring and semi-natural-language authoring.
In snippet-based authoring, 
developers specify characteristics of elements to match by writing partial code snippets. 
In semi-natural language authoring, 
a textual representation offers a representation for understanding design rules and resolving ambiguities. 
We implemented these approaches in RulePad.
To evaluate RulePad, we conducted a between-subjects  study with 14 participants comparing RulePad to the PMD Designer, a utility for writing rules in a popular rule checker. We found that those with RulePad were able to successfully author 13 times more query elements in significantly less time and reported being significantly more willing to use RulePad in their everyday work. 
\end{abstract}




\maketitle

\newcommand{\GUI}{\textsc{Graphical Editor}}
\newcommand{\TE}{\textsc{Textual Editor}}
\newcommand{\ind}{\indent\indent}

\definecolor{quantColor}{rgb}{0.93, 0.35, 0.0}
\definecolor{constrColor}{rgb}{0.55, 0.0, 0.55}

\definecolor{purpleDiagram}{rgb}{0.58, 0.0, 0.83}
\definecolor{greenDiagram}{RGB}{69,172,15} 
\definecolor{brownDiagram}{rgb}{0.59, 0.29, 0.0}
	
\def\checkmark{\tikz\fill[scale=0.4](0,.35) -- (.25,0) -- (1,.7) -- (.25,.15) -- cycle;} 
\newcommand{\cmark}{\ding{51}}%
\newcommand{\xmark}{\ding{55}}%

\newcommand{\code}[1] {{\smaller\texttt{#1}}}

\newenvironment{quoteRule}%
  {\list{}{\leftmargin=0.2in\rightmargin=0.1in}\item[]}%
  {\endlist}

\section{introduction}





Writing good documentation has long been viewed as key to helping developers successfully follow and understand the rationale behind design decisions, helping to prevent software defects and code decay~\cite{parnas1986rational}. 
A design decision encompasses rationale explaining developers' reasoning which led to the decision as well as constraints imposed by the design decision on how code can be written. In this paper, we refer to the constraints imposed by a design decision as a {\it design rule}~\cite{baldwin2000design}. 
Traditional approaches to documentation rely on developers first writing decisions down in documents or comments and then consistently updating the documentation as the code changes. 
However, this often does not occur in practice~\cite{lethbridge2003software}, resulting in documentation which is inaccurate, incomplete, and untrustworthy~\cite{headICSE17}, and encouraging developers to ignore documentation and reverse engineer design directly from code~\cite{latoza2006maintaining}. This makes questions about design rationale some of the most frequently reported hard-to-answer questions~\cite{latoza2010hard} and one of the most serious problems developers report facing~\cite{latoza2006maintaining}.


A potential solution is to ensure that design rules are \textit{checkable}, expressed in a representation where a program analysis tool can continuously check for inconsistencies and flag divergences between the documented rule and code \cite{Mehrpour2019ActiveDoc}.   
To achieve this, developers might use static analysis tools such as CheckStyle~\cite{burn2005checkstyle}, PMD~\cite{copeland2005pmd}, and FindBugs~\cite{hovemeyer2004findbugs} to write custom, project-specific rules describing a defect pattern to avoid. 
However, 
existing tools require developers to either write program analyses in a general purpose programming language (e.g. FindBugs~\cite{hovemeyer2004findbugs})
or use complex query notations to describe patterns (e.g. XPath in PMD~\cite{copeland2005pmd}).
This restricts authoring and changing design rules to those with specialized program analysis knowledge, preventing most developers from creating or maintaining design knowledge through checkable design rules. 








To author checkable design rules, developers should be able to work with design rules in simple and expressive representations. 
We propose two techniques for authoring checkable design rules: \textit{snippet-based authoring} in which design rules are represented in code-based templates and \textit{semi-natural language authoring}. These techniques are complimentary,
offering different levels of simplicity and expressiveness.
Snippet-based authoring enables developers to easily express rule through templates that look like code, but may be ambiguous with complex rules. 
Semi-natural language authoring lets developers author design rules in an expressive and unambiguous textual representation, but while simpler than complex query notations, requires the developer to learn a semi-natural language.

We implemented these techniques in a system for authoring checkable design rules, RulePad, as a 
design rule {\GUI} and {\TE}. 
As developers edit design rules in the {\GUI}, they receive immediate feedback and can view examples of code which satisfies and violates the rule. As a rule is constructed in the {\GUI}, RulePad constructs a natural language textual representation of the rule using the grammar of the {\TE}, which is bidirectionally synchronized. Developers may also edit or author new rules using the textual representation, including resolving ambiguities in situations where the 
graphical representation of a rule is ambiguous (e.g., elements are combined through either conjunction or disjunction). 

To evaluate the ability of RulePad to enable developers to author checkable design rules more easily, we conducted a user study. 14 participants were asked to author design rules with either RulePad or PMD, an existing widely-used static rule checker. This study also offered a first usability evaluation of using PMD to author custom rules.
Participants with RulePad were able to successfully author 13 times more query elements in significantly less time and reported being significantly more willing to use RulePad in their everyday work. 
Participants working with the PMD Designer experienced challenges with both the tool and its documentation which hindered them from making significant progress.

In the following sections, we first review design rules, present a motivating example, describe the two approaches for authoring design rules, describe the design of our system, and report the results from our evaluation study. We conclude with related work and a discussion of potential future directions.

\section{Design Rules}\label{sec:design_rules}

When writing code, developers make design decisions choosing between alternatives, formulating design rules specifying constraints to ensure that code is consistent with the design. Violating these constraints causes code to decay and drift from the originally intended design, which may alter the code's behavior, reduce its maintainability, or prevent extensibility. 

Design rules vary in their scope and complexity. Some may manifest as statement-level constraints. For example, consider an application in which data is persisted. To ensure correct retrieval of persisted data in concurrently executing code, a design rule might impose immediate persistence, requiring a specific method call to immediately save data:
\begin{tcolorbox}[boxsep=2pt,left=2pt,right=2pt,top=2pt,bottom=2pt,colback=gray!10,arc=0pt,outer arc=2pt,colframe=gray!50]
\small
\code{Save()} calls should always be committed immediately. \\
{\color{quantColor} \textit{IF} the \code{save()} method of the persistence library is called} 
{\color{constrColor} \textit{THEN} the \code{now()} method must be followed immediately.}
\begin{tcolorbox}[boxsep=2pt,left=2pt,right=2pt,top=2pt,bottom=2pt,colback=white,colframe=gray!50]
\code{{\color{quantColor} ofy().save()}.entity(this){\color{constrColor}.now()};}
\end{tcolorbox}
\end{tcolorbox}

Design rules may also manifest as constraints on how classes or methods are declared. For example, in order for specific data to be persisted as expected, a design rule might dictate that all subclasses must be persisted by requiring a specific annotation:
\begin{tcolorbox}[boxsep=2pt,left=2pt,right=2pt,top=2pt,bottom=2pt,colback=gray!10,arc=0pt,outer arc=2pt,colframe=gray!50]
\small
Artifacts should be marked for persistence with Entity annotations.\\
{\color{quantColor}\textit{IF} an object is an artifact subclass} 
{\color{constrColor}\textit{THEN} it needs to be an entity marked with \code{@Subclass}}.
\begin{tcolorbox}[boxsep=2pt,left=2pt,right=2pt,top=2pt,bottom=2pt,colback=white,colframe=gray!50]
\code{{\color{constrColor}@Subclass(index=true)}\\
public class ADT {\color{quantColor}extends Artifact} \{ ... \}
}
\end{tcolorbox}
\end{tcolorbox}

Design rules may also express higher-level constraints, such as an architectural style. High-level design rules may be decomposed into lower-level design rules that can then be directly checked.
For example, adopting a sharded architectural style might be expressed through several rules. One might specify a constraint to create intermediate classes (a Command) to capture communication between shards and a second might specify how sharded commands are executed.


\begin{tcolorbox}[boxsep=2pt,left=2pt,right=2pt,top=2pt,bottom=2pt,colback=gray!10,arc=0pt,outer arc=2pt,colframe=gray!50]
\small
All microtask commands must be handled by \code{Command} subclasses.\\
{\color{quantColor} \textit{IF} a method is a \code{static} method on Command}
{\color{constrColor} \textit{THEN} it should implement its behavior by constructing a new \code{Command} subclass instance}. 
\begin{tcolorbox}[boxsep=2pt,left=2pt,right=2pt,top=2pt,bottom=2pt,colback=white,colframe=gray!50]
\code{public abstract class FunctionCommand {\color{quantColor}extends Command} \{ \\
public {\color{quantColor}static} FunctionCommand create(...) \{ \\
		return {\color{constrColor}new Create(...)}; 
	\} \\
protected static class {\color{constrColor}Create extends FunctionCommand} \{ .. \}\\
... \}
}
\end{tcolorbox}
\end{tcolorbox}


\begin{tcolorbox}[boxsep=2pt,left=2pt,right=2pt,top=2pt,bottom=2pt,colback=gray!10,arc=0pt,outer arc=2pt,colframe=gray!50]
\small
\code{Commands} must implement \textit{execute}.\\
{\color{quantColor}\textit{IF} a class is a subclass of \code{Command}}
{\color{constrColor} \textit{THEN} it must implement execute.}
\begin{tcolorbox}[boxsep=2pt,left=2pt,right=2pt,top=2pt,bottom=2pt,colback=white,colframe=gray!50]
\code{public abstract class {\color{quantColor}ADTCommand extends Command} \{ \\
protected static class Create {\color{quantColor} extends ADTCommand} \{ \\
public void {\color{constrColor} execute}(Function funct, String projectId) \{ \\
... \} ... \}
}
\end{tcolorbox}
\end{tcolorbox}

Design rules can be expressed through an IF/THEN
structure.
This structure specifies {\it when} and {\it how} it should apply. 
This can be expressed through a \textit{quantifier} describing when the rule applies and \textit{constraints} describing what must be true. 

\section{Motivating Examples}\label{sec:user_experience}
\begin{figure}[t]
    \centering
    \includegraphics[width=.47\textwidth]{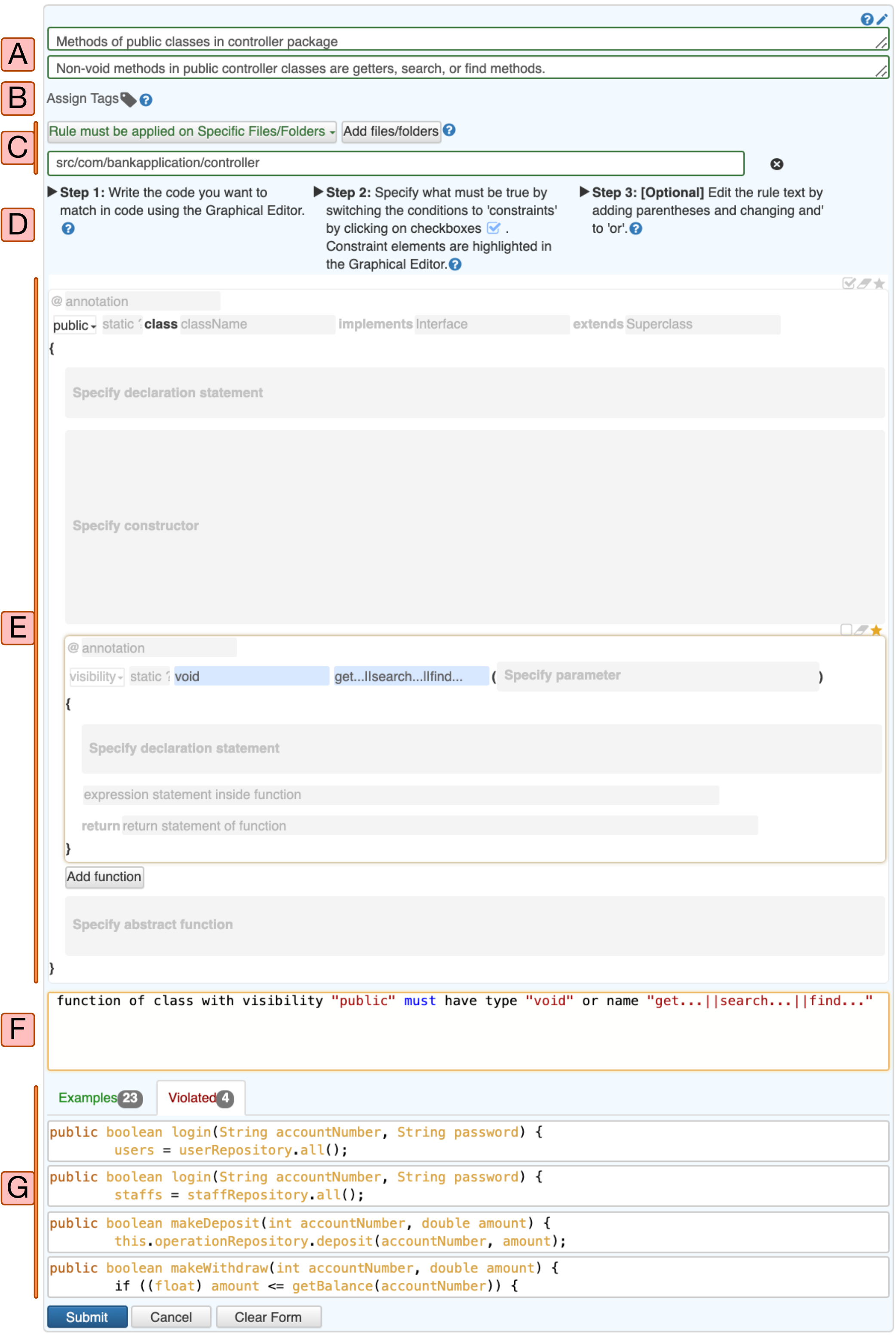}
    \caption{A developer using RulePad to create a new design rule may first specify a title and description (A), assign tags (B), and specify where the rule applies (C). 
    Using the {\GUI} (E), the developer may then write code snippets in a structured editor, interactively refining the rule to resolve ambiguity. Developers may verify the rule by (F) reading a description in the {\TE}, which they may also optionally edit. While authoring rules, developers may (G) view a list of code snippets which satisfy and violate the rule to check its behavior.
    }
    \label{fig:edit_rule}
\end{figure}


We illustrate how RulePad enables developers to author design rules through three examples. 


\subsection{Basic Authoring}
Alice is a developer working on a bank  application\footnote{https://github.com/derickfelix/BankApplication}. While implementing a feature, she formulates a design rule about the responsibilities of classes in the \texttt{\small Controller} package by expressing a constraint on the allowed public methods. 


\begin{quote}
    If a method is in a \texttt{\small public} class, then it must be \texttt{\small void} or have a name which begins with \texttt{\small get}, \texttt{\small search}, or \texttt{\small find}.
\end{quote}

Alice opens 
\textit{RulePad} and reads a brief tutorial explaining each element of the interface and offering an example of how to write a simple design rule. 
She writes a title and a description explaining the rationale, assigns tags, and specifies that the rule should apply only to files in the \texttt{\small controller} package (Figure~\ref{fig:edit_rule}.A-C).
Alice notices the dynamic guide (Figure~\ref{fig:edit_rule}.D), which lists three steps, with the first highlighted.
The first step instructs her to describe the code to matched in the {\GUI}. 
Alice writes the IF part of the rule, \texttt{\small public} class, specifying \texttt{\small public} in the dropdown (Figure~\ref{fig:edit_rule}.E). 

The dynamic guide now indicates that the first step has been finished and highlights the second step. The second step asks her to specify the constraints which matched code must satisfy by marking one or more of the elements as constraints. She specifies that the function return type should be \texttt{\small void}. Realizing she needs to match function names which start with a prefix, 
she writes \texttt{\small get...$\mid\mid$search...$\mid\mid$find...} to match identifiers which begin with \texttt{\small get}, \texttt{\small search}, or \texttt{\small find}. She then checks the boxes to indicate that both the function return type and identifier should be treated as constraints.


Returning to the dynamic guide, the third step suggests she consider editing the rule in the {\TE} to change `and' to `or', if necessary. She reads the generated text:   \texttt{\small class with visibility "public" must have function with (type "void" and name "get...$\mid\mid$ search...$\mid\mid$find..."}).
Realizing this is not what she intended, she edits `and' to `or' (Figure~\ref{fig:edit_rule}.F). 

To check if the rule behaves as she expects, Alice inspects the list of matches and violations (Figure~\ref{fig:edit_rule}.G). She sees that the rule is still not quite as she intended. Two additional method identifiers should be included: \texttt{\small login} and \texttt{\small make...}. She edits the method name pattern matching expression in the {\GUI}. Glancing again at the updated list of matches and violations, she finds that the rule now works as she intends.

\subsection{Advanced Authoring}\label{sec:adv_authoring}

Working on another feature, Alice formulates a rule: 
\begin{quote}
    If a class is \texttt{\small public}, then it must have at least one method whose name begins with \texttt{\small get}.
\end{quote}

\begin{figure}[t]
    \centering
\includegraphics[width=.45\textwidth]{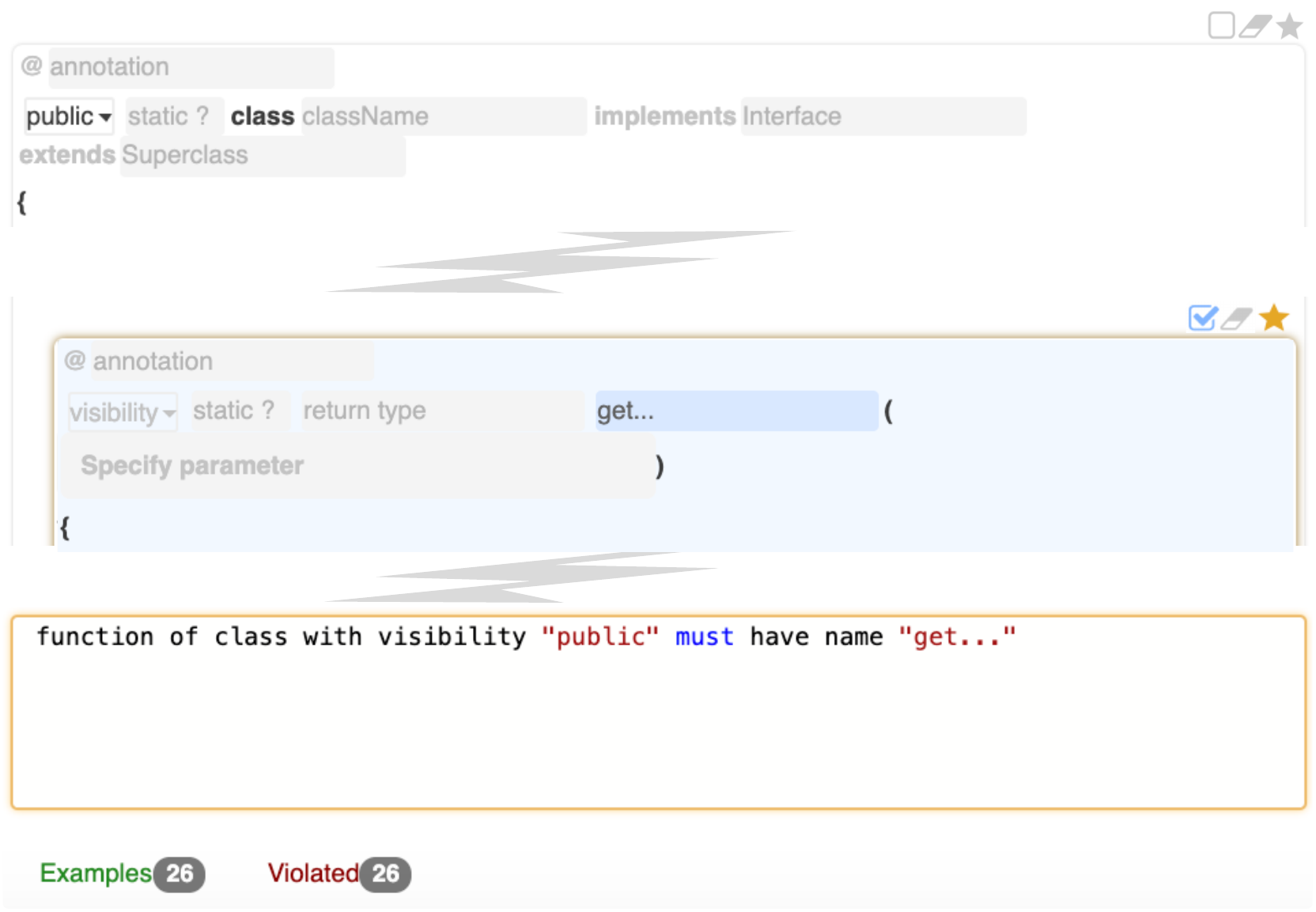}
    \caption{By default, RulePad use the lowest common ancestor of constraint elements as the Element of Interest. In this example, there is a single constraint element, "get...". RulePad selects the containing element, a method, as the Element of Interest, indicated by a golden star. 
    }
    \label{fig:eoi_example_1}
\end{figure}

\begin{figure}[t]
    \centering
\includegraphics[width=.45\textwidth]{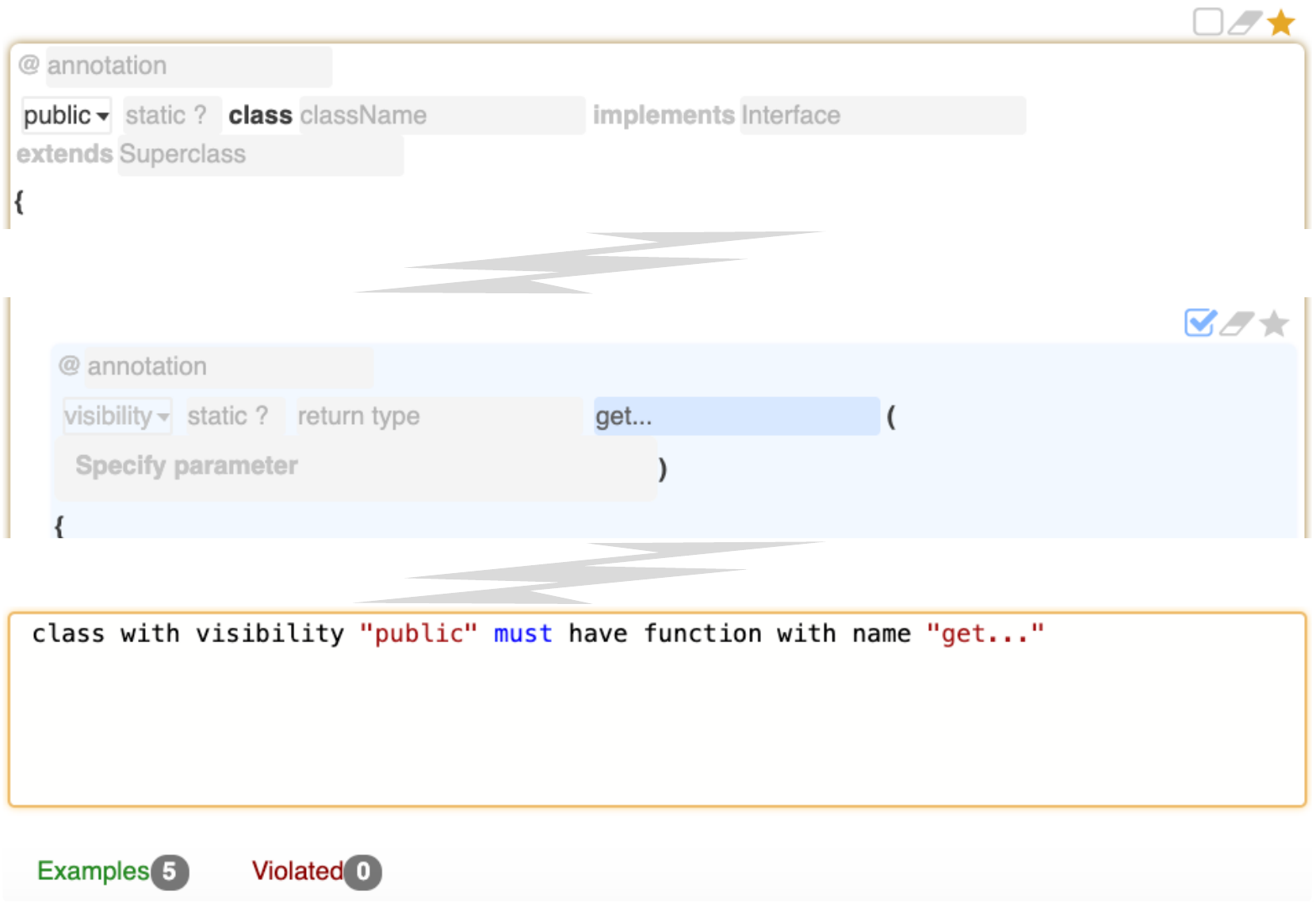}
    \caption{Clicking the golden star on the class updates the Element of Interest to the class. This is reflected in the updated textual description and updated list of violations.}
    \label{fig:eoi_example_2}
\end{figure}

While creating the design rule, 
she notices that the {\TE} indicates that it is matching the {\it methods} in \texttt{\small public} classes rather than the \texttt{\small public} {\it classes} themselves. The tool indicates that there are 26 violations (Figure~\ref{fig:eoi_example_1}), which should not be violations. Recalling the tutorial, she realizes that she needs to set the class to be the element of interest. She clicks on the star on the right side of the class element. The text in the {\TE} now indicates that \texttt{\small public} classes will be matched~ (Figure~\ref{fig:eoi_example_2}). 

\subsection{Authoring Design Rules through the Textual Editor}

Alice formulates a rule that

\begin{quote}
    If there is a field in a class, then it must be \texttt{\small private}.
\end{quote}

\begin{figure}[t]
    \centering
\includegraphics[width=.45\textwidth]{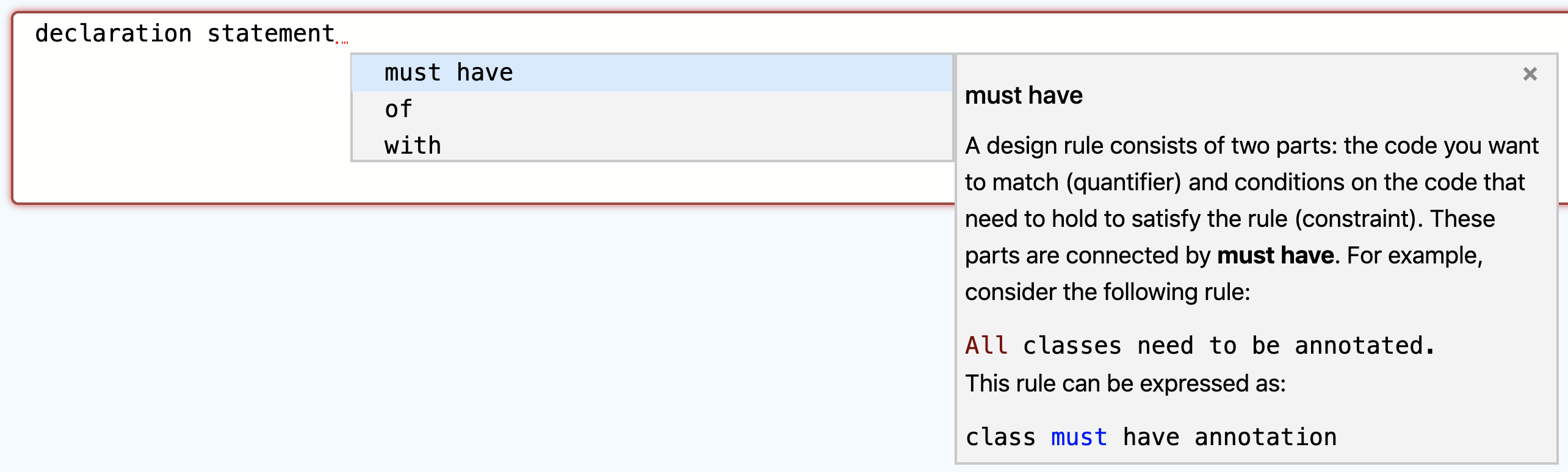}
    \caption{Developers can write design rules using the {\TE}.  
    Autocomplete provides suggestions,  including documentation and examples of its use.}
    \label{fig:auto_complete}
\end{figure}

After seeing the representation of the past rules she has authored in the {\TE}, she decides to write the rule using the {\TE}. Based on her prior experience, she begins the design rule with the Element of Interest, \texttt{\small declaration statements}. As she types, autocomplete provides suggestions (Figure~\ref{fig:auto_complete}). Alice reads the documentation of each suggestion and selects the next token by clicking on the suggestion. 

\section{Techniques For Authoring Design Rules}


Developers traditionally document design rules in design documents, such as by using example code snippets describing how to write code correctly or in prose describing constraints on how code should be written. RulePad enables developers to author and work with design rules in these forms.

A key challenge in supporting authoring design rules is the inherent tradeoff between simplicity and expressiveness in representing rules. We introduce snippet-based authoring and semi-natural language authoring as complimentary techniques for authoring checkable design rules. 
Representing the design rule in a representation close to what will be matched ---code--- enables a direct manipulation interaction, where developers can rapidly write and edit what they wish to see matched and immediately see what is matched.
Using the {\GUI}, developers can then begin to refine rules in increasingly complex ways, using code-snippets to express the essence of the rule and interactively refining it as necessary. 
In more complex rules, additional information is needed to resolve ambiguities. The {\TE} enables developers to understand how the code snippets they have added are composed to form a design rule as well as enabling more experienced developers more control. Our key goal was to keep the design of 
of RulePad simple enough that developers inexperienced with program analysis and query notations are able to rapidly author design rules while making it expressive enough to cover most of the design rules expressible in existing AST-based rule checkers.



\subsection{Snippet-Based Authoring}\label{sec:snippetBasedAuthoring}


Enabling developers to author design rules using code snippets is a natural and appealing approach. Developers already know code, and can simply write down what they wish to match. However, code snippets by themselves are often ambiguous. Consider:
\begin{quote}
\small
\texttt{"Mapper     String"}
\end{quote}
Mapper seems to be a name, but for what? And how is String related? Additional information is needed to determine what AST elements these code snippets should be matched against and how these elements are related. 
Using code-snippets entered in a traditional text editor is not straightforward, as the text is ambiguous. 

To address this challenge, we introduce rule authoring through code-based templates.
The code-based template enables developers to specify this information using a structured editing experience to select options and type text into boxes which correspond to specific AST elements that the developer wishes to match.


Other additional information is sometimes needed. 
Elements may define a quantifier or constraints of design rules (Section~\ref{sec:design_rules}), describing either when the rule applies or what code must exist when it does apply. Design rules must also specify an Element of Interest (EoI) describing to which element it applies. Without specifying this, the rule may be ambiguous. For example, while the design rules in Figures~\ref{fig:eoi_example_1} and \ref{fig:eoi_example_2} have identical partial code snippets, they have different meanings. The design rule in Figure \ref{fig:eoi_example_1} applies to functions, indicating that all public functions must be getters. The design rule in Figure~\ref{fig:eoi_example_2} instead applies to classes, indicating that all public classes must have getters. To resolve this ambiguity, developers can interactively indicate the element of interest in the interface.



\subsection{Semi-Natural-Language Authoring}\label{sec:semiNaturalLangAuthoring}

\renewcommand{\arraystretch}{1.1}
\begin{table*}[t]
\centering
\small
\caption{The semi-natural language offers a more natural and compact representation of design rules. The top row lists a design rule and its corresponding representation in the semi-natural language (second row) and a traditional query language, XPath (bottom row).}
\begin{tabular}{p{.98\textwidth}}
\rowcolor{gray!20} Model classes should have `private' fields and getters. \\ \hline
\texttt{\small class must have declaration statement with visibility "private" and function with name "get..."}  \\\hline
\texttt{\small//CompilationUnit[PackageDeclaration/Name[@Image="com.bankapplication.model"]]//ClassOrInterfaceDeclaration[count(
ClassOrInterfaceBody/ClassOrInterfaceBodyDeclaration/FieldDeclaration[@Private="true"])=0 or count(ClassOrInterfaceBody/ ClassOrInterfaceBodyDeclaration/MethodDeclaration/MethodDeclarator[starts-with(@Image,"get")])=0]} \\ \hline  
\end{tabular}
\label{tab:XPath_example}
\end{table*}

Developers often author design rules in text, as it can contain explanations and other information about design rules. However, natural language text is difficult to use directly by tools, particularly as it may be ambiguous. 

Traditionally, rule checkers have met this challenge by introducing specialized query notations,
such as XPath queries. However, these notations are often unnatural, with little resemblance to natural language. This imposes a barrier to both rule authors and readers. To understand rules written in these notations, developers first need to know the syntax and meaning of the notation. 
To author rules, users must learn how to express the desired rule correctly in the notation. 
In particular, developers must learn how specific AST elements map into a corresponding representation in the query notation.


To make it easier for developers to work with design rules in textual form, we created a semi-natural language for design rules.
Our goal was to create a language with easily understandable syntax and semantics, simplifying the complexity of the underlying mapping to the complex notations used by analysis tools. 
The language is translated into XPath queries by RulePad, but masks the complexity of this notation from the developer. 

Design rules are expressed using an IF/THEN structure, with a quantifier followed by a constraint. In the design rule syntax, this take the form of text which first describes the quantifier, followed by `must have', and then text describing one or more constraints. Rather than use an IF/THEN syntax, we chose to encode the IF/THEN structure using a must have syntax, as it is more compact.  
For example, compare
`\texttt{\small class with visibility "private" must have function with name "get..."}' 
and
`\texttt{\small IF class has visibility "private" THEN it must have function with name "get..."}'.
The must have syntax requires 10 words and only one reference to `class', while in the IF/THEN syntax, there are 13 words and two references to `class', one potentially ambiguous.

There are 6 main rules in the grammar.

\begin{enumerate}[itemsep=2pt,parsep=2pt]
    \item Values to match in the code are specified through double-quoted text, with pattern matching allowed.
    \begin{quote}
    \small
        \texttt{\textbf{"...Mapper"}}
    \end{quote}
    \item Conditions imposed on an element are listed after the element. AST children of an element may be specified using \textbf{`with'} and an AST parent specified using \textbf{`of'}. 
    \begin{quote}
    \small
        \texttt{function \textbf{with} parameter \textbf{of} class }
    \end{quote}
    \item Quoted values are not prefixed with `with', except for implementation and extension which are prefixed with `of'.
    \begin{quote}
    \small
        \texttt{\textbf{name "...Mapper"}}\\
        \texttt{\textbf{implementation of "I..."}}\\
        \texttt{\textbf{extension of "...Repository"}}
    \end{quote}
    \item The quantifier is specified first, followed by \textbf{`must have'}, followed by the constraints.
    \begin{quote}
    \small
        \texttt{function with type "void" of class \textbf{must have} name "update||destroy"}
    \end{quote}
    \item Multiple AST children may be specified using \textbf{`and'} and \textbf{`or'}.
    \begin{quote}
    \small
        \texttt{function must have name "...Mapper" \textbf{and} visibility "public" \textbf{or} type "void"}
    \end{quote}
    \item Parentheses can be used around AST children to resolve ambiguity.
    \begin{quote}
    \small
        \texttt{function must have \textbf{((}name "...Mapper" and visibility "public"\textbf{)} or type "void"\textbf{)}}
    \end{quote}
\end{enumerate}

\section{RulePad}\label{sec:system}

\begin{figure}[t]
\centering
\includegraphics[width=.48\textwidth]{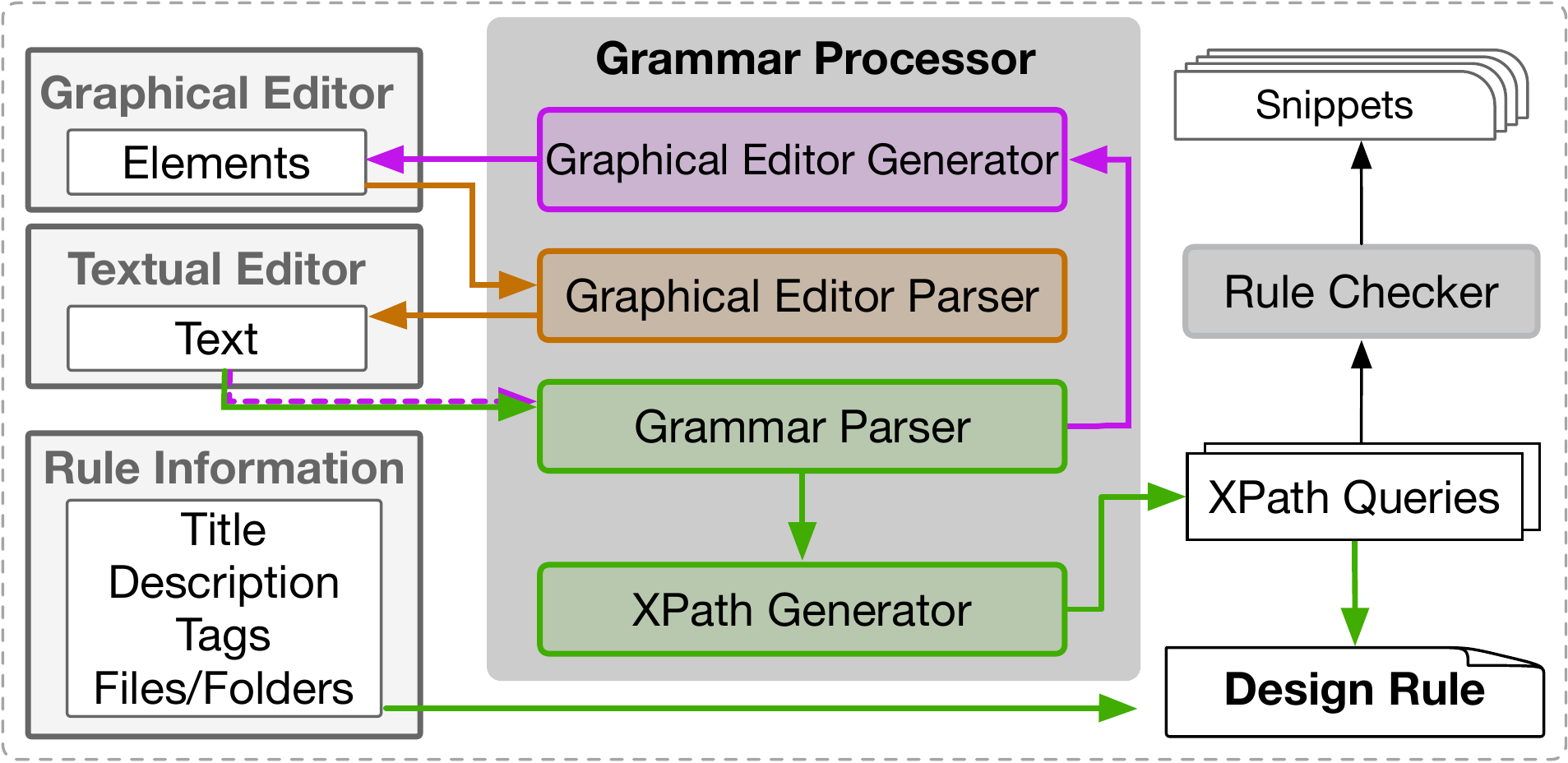}
\caption{The grammar processor bidirectionally synchronizes the {\GUI} with the {\TE} and enables checking rules against code through XPath.}
\label{fig:rulePad_system}
\end{figure}

RulePad enables developers to work with design rules through a {\TE} and {\GUI}.
RulePad integrates these into a unified interface for authoring design rules, built on top of the ActiveDocumentation system for working with design rules~\cite{Mehrpour2019ActiveDoc}.  

The {\TE} and the {\GUI} are bidirectionally synchronized, enabling developers to switch back and forth as they incrementally construct a design rule (Figure~\ref{fig:rulePad_system}). Over time, this may facilitate a developer incrementally learning how to express specific rules in the semi-natural language.
%
Code written in the {\GUI} is processed by a graphical editor parser to generate a textual representation according to the grammar ({\color{brownDiagram}\textbf{brown}} path in Figure~\ref{fig:rulePad_system}).
If the text in the {\TE} is changed, the grammar parser, built on the ANTLR grammar parser~\cite{parr2013ANTLR}, generates a parse tree.
The parse tree is then traversed to create code in the {\GUI} ({\color{purpleDiagram}\textbf{purple}} path in Figure~\ref{fig:rulePad_system}).

In RulePad, design rules are checked against the code for conformance.
XPath queries are first generated by traversing the parse tree and mapping each node to an XPath element ({\color{greenDiagram}\textbf{green}} path in Figure~\ref{fig:rulePad_system}). For example, the "\textit{child} \textbf{of} \textit{parent}" element in the grammar is mapped to \texttt{\smaller"parent/child"} in XPath, and "\textit{element} \textbf{with} \textit{conditions}" is mapped to \texttt{\smaller"element[conditions]"} in XPath. The \textbf{must have} construct enables creating two XPath queries, the quantifier and the constraints queries. 
As in many traditional rule checkers,
RulePad checks the XPath queries against code through a custom rule checker. Each design rule is translated into two separate XPath queries, corresponding to the quantifier and constraint. By comparing the results from each query, RulePad is able to identify code snippets that satisfy the rule as well as violations.  

\subsection{The {\GUI}}
The {\GUI} offers developers a design rule template to begin authoring, uses progressive disclosure to reduce visual clutter, supports pattern matching of identifiers through pattern matching expressions, supports specifying an element of interest, and supports labeling conditions as a quantifier or constraint.

\subsubsection{Design Rule Template}
Developers interact with design rules in the {\GUI} through a code-based template which displays an outline of a class declaration as the main element.
Within the template, common characteristics for each element are displayed, which the user may choose to specify in the design rule. For example, for a class, its annotations, visibility, specifier (e.g. {\it static}), 
name, superclass, and interfaces it implements are displayed. 
Contained within the template of the class are additional elements which may also be specified and added to the rule: fields, constructors, methods, and abstract methods.
Additional instances of an element may be added to the rule using the ``Add [name of the element]'' button. 

\subsubsection{Progressive Disclosure}

To reduce clutter and distraction, the {\GUI} employs three levels of progressive disclosure, graying out text until needed. Elements may be inactive, of potential interest, or active. 
Inactive elements have no specified characteristics and have no role in the rule. As they are not yet relevant, they are indicated with a lighter font color or overlaid shading.
When the user hovers over an inactive element, the element becomes of potential interest and the visibility of its color is increased.
After elements are given at least one specified characteristic, they become active and are incorporated into the design rule. Active elements are highlighted with a darker font  (Figure~\ref{fig:edit_rule}.E).

\subsubsection{Pattern Matching}\label{sec:pattern_matching}

The {\GUI} supports pattern matching expressions with a notation that is code-like and may include simple wild cards. These expressions enable users to write patterns that match textual values including \textit{ending with a sub-string}, \textit{starting with a sub-string}, \textit{containing a sub-string}, their negations, and combinations of patterns. To avoid confusion with the dereference operator in C++, three dots are used to denote \textit{any sequence of characters}. For example, \texttt{\small ...Repository} matches names ending with \texttt{\small Repository}, 
\texttt{\small !BaseRepository} matches names that are not \texttt{\small BaseRepository},
and \texttt{\small !BaseRepository\&\&...Repository} matches names that are not equal to \texttt{\small BaseRepository} and end with \texttt{\small Repository}.

\subsubsection{Constraints}
Constraints specify what must be true about the quantifier to satisfy the rule. In the {\GUI}, both the quantifier and the constraints are displayed in a single view. 
By default, elements are set as a quantifier when added. This enables users to then incrementally add constraints by using checkboxes to toggle elements from a quantifier to a constraint. If an element is selected as a constraint, all of its properties (the children of its AST node) are also marked as constraints. Constraints are indicated through a blue background color (Figure~\ref{fig:edit_rule}.E). 

\subsubsection{Dynamic Guide}
In evaluations of early prototypes, we found that users sometimes got stuck with how to make progress in authoring a design rule, particularly in initially discovering that each design rule must include a constraint. To help users understand where to start, the dynamic guide walks developers through the steps of authoring a design rule.
As the developer completes each step, the next is highlighted.
The user is prompted to write a quantifier, specify a constraint, and optionally edit the logical connectives between conditions.

\subsubsection{Elements of Interest}
A default EoI is generated by computing the \textit{lowest common ancestor} of all constraints, indicated through a golden star. For example, if the name of a function and its parameter type are both specified and marked as constraints, the function declaration is selected as the EoI for the rule.
Elements which may be selected as the EoI are indicated through a grey star.  

\subsection{The {\TE}}
The {\TE} enables developers to author and edit design rules in a semi-natural language.
To help developers begin to understand and use the grammar, RulePad offers developers autocomplete suggestions. To help users understand the meaning of each potential completion, each is accompanied by context-specific documentation. In addition, developers may investigate the meaning of a token in the {\TE} by following a visual link to the equivalent element in the {\GUI}.

\renewcommand{\arraystretch}{1.1}
\begin{table}[t]
\centering
\setlength\tabcolsep{.7pt}
\caption{The semi-natural language grammar. Bold indicates terminal nodes, italic non-terminals, brackets optional nodes, and curly brackets zero or more occurrences. \textit{symbol} specifies valid Java symbols such as parentheses and dot, and \textit{character} specifies valid characters for Java identifiers. }
\begin{tabular}{lr p{.38\textwidth}}

{\smaller 1} & \textit{DR} ::= &
    \textit{c} \textbf{\texttt{\smaller must have}} \textit{cExp}
    | \textit{m} \textbf{\texttt{\smaller must have}} \textit{mExp} 
   | \newline 
   \textit{am} \textbf{\texttt{\smaller must have}} \textit{amExp}  
   | \textit{co} \textbf{\texttt{\smaller must have}} \textit{coExp} 
   | \newline
    \textit{d} \textbf{\texttt{\smaller must have}} \textit{dExp} 
    | \textit{p} \textbf{\texttt{\smaller must have}} \textit{pExp} \\
   
{\smaller 2} & \textit{c}  ::= & \textbf{\texttt{\smaller class}} [\textbf{\texttt{\smaller with}} \textit{cExp}] [\textbf{\texttt{\smaller of}} \textit{c}]\\   
{\smaller 3} & \textit{m}  ::= & \textbf{\texttt{\smaller function}} [\textbf{\texttt{\smaller with}} \textit{mExp}] [\textbf{\texttt{\smaller of}} \textit{c}]\\
{\smaller 4} & \textit{am}  ::= & \textbf{\texttt{\smaller abstract function}} [\textbf{\texttt{\smaller with}} \textit{amExp}] [\textbf{\texttt{\smaller of}} \textit{c}]\\
{\smaller 5} & \textit{co}  ::= & \textbf{\texttt{\smaller constructor}} [\textbf{\texttt{\smaller with}} \textit{coExp}] [\textbf{\texttt{\smaller of}} \textit{c}]\\   
{\smaller 6} & \textit{d}  ::= & \textbf{\texttt{\smaller declaration statement}} [\textbf{\texttt{\smaller with}} \textit{dExp}] [\textbf{\texttt{\smaller of}} (\textit{c} | \textit{m} | \textit{co})]\\
{\smaller 7} & \textit{p}  ::= & \textbf{\texttt{\smaller parameter}} [\textbf{\texttt{\smaller with}} \textit{pExp}]\\
{\smaller 8} & \textit{t}  ::= & \textbf{\texttt{\smaller type}} [\textit{pattern} | \textit{expr}]\\
{\smaller 9} & \textit{e}  ::= & \textbf{\texttt{\smaller extension of}} (\textit{pattern} | \textbf{\texttt{\smaller superclass}})\\
{\smaller 10} & \textit{im}  ::= & \textbf{\texttt{\smaller implementation of}} (\textit{pattern} | \textbf{\texttt{\smaller interface}})\\
{\smaller 11} & \textit{ex}  ::= & \textbf{\texttt{\smaller expression statement}} [\textit{expr}] [\textbf{\texttt{\smaller of}} (\textit{m} | \textit{co})]\\
{\smaller 12} & \textit{i}  ::= & \textbf{\texttt{\smaller initial value}} [\textit{expr}] [\textbf{\texttt{\smaller of}} \textit{d}]\\   
{\smaller 13} & \textit{r}  ::= & \textbf{\texttt{\smaller return value}} [\textit{expr}]\\
{\smaller 14} & \textit{a}  ::= & \textbf{\texttt{\smaller annotation}} [\textit{expr}]\\
{\smaller 15} & \textit{n}  ::= & \textbf{\texttt{\smaller name}} [\textit{pattern}]\\
{\smaller 16} & \textit{s}  ::= & \textbf{\texttt{\smaller specifier}} [\textit{pattern}]\\
{\smaller 17} & \textit{v}  ::= & \textbf{\texttt{\smaller visibility}} [\textit{pattern}]\\

&\\ [-2ex]

{\smaller 18} & \textit{cExp} ::= &
      \textbf{\texttt{\smaller (}}\textit{cExp}\textbf{\texttt{\smaller )}}  | \textit{cExp} \textit{op} \textit{cExp}  | \textit{cExp} | \newline
      \textit{a} | \textit{s} | \textit{v} | \textit{n} | \textit{e} | \textit{im} | \textit{m} | \textit{am} | \textit{co} | \textit{d} | \textit{c} | \textit{r}
\\
{\smaller 19} & \textit{mExp} ::= &
      \textbf{\texttt{\smaller (}}\textit{mExp}\textbf{\texttt{\smaller )}}  | \textit{mExp} \textit{op}  \textit{mExp}  | \textit{mExp} | \newline
      \textit{a} | \textit{s} | \textit{v} | \textit{t} | \textit{n} | \textit{p} | \textit{r} | \textit{d} | \textit{ex}
\\
{\smaller 20} & \textit{amExp} ::= &
      \textbf{\texttt{\smaller (}}\textit{amExp}\textbf{\texttt{\smaller )}} | \textit{amExp} \textit{op} \textit{amExp} | \textit{amExp} | 
      \textit{a} | \textit{s} | \textit{v} | \textit{t} | \textit{n} | \textit{p}
\\
{\smaller 21} & \textit{coExp} ::= &
      \textbf{\texttt{\smaller (}}\textit{coExp}\textbf{\texttt{\smaller )}}  | \textit{coExp} \textit{op} \textit{coExp} | \textit{coExp} | 
      \textit{a} | \textit{s} | \textit{v} | \textit{p} | \textit{r} | \textit{d} | \textit{ex}
\\
{\smaller 22} & \textit{dExp} ::= &
      \textbf{\texttt{\smaller (}}\textit{dExp}\textbf{\texttt{\smaller )}}  | \textit{dExp} \textit{op} \textit{dExp} | \textit{dExp} | 
      \textit{a} | \textit{s} | \textit{v} | \textit{t} | \textit{n} | \textit{i}
\\
{\smaller 23} & \textit{pExp} ::= &
      \textbf{\texttt{\smaller (}}\textit{pExp}\textbf{\texttt{\smaller )}}  | \textit{pExp} \textit{op} \textit{pExp} | \textit{pExp} | 
      \textit{t} | \textit{n}
\\

&\\ [-2ex]
{\smaller 24} & \textit{op} ::= & \textbf{\texttt{\smaller and}} | \textbf{\texttt{\smaller or}} \\
{\smaller 25} & \textit{pattern} ::= &  \textbf{\texttt{\smaller"}} \{ \textit{part} ( \textbf{\texttt{\smaller \&\&}} |  \textbf{\texttt{\smaller ||}} ) \} \textit{part}\textbf{\texttt{\smaller "}} \\

{\smaller 26} & \textit{part} ::= & [\textbf{\texttt{\smaller!}}][\textbf{\texttt{\large...}}] \textit{character}\{\textit{character}\} [\textbf{\texttt{\large...}}] \\

{\smaller 27} & \textit{expr}   ::=  & \textbf{\texttt{\smaller "}} (\textit{character} | \textit{symbol}) \{\textit{character} | \textit{symbol}\} \textbf{\texttt{\smaller"}} \\

\end{tabular}
\label{tab:grammar}
\end{table}

\subsubsection{Grammar}\label{sec:grammar}

The semi-natural language is defined by a grammar, implemented using the ANTLR parser generator~\cite{parr2013ANTLR} (Table~\ref{tab:grammar}). 
The grammar currently supports 16 AST element types (rules~2-17), including class declarations, method and constructor declarations, declaration statements, and expression statements. 
Using the grammar, design rules are created with a structural construct of an AST element followed by \textbf{\texttt{\smaller must have}} and the expression construct of the same AST element (rule~1). 
The structural construct of an AST element starts with a terminal node (i.e., name of the element) which is followed by expression constructs, as children, ([\textbf{\texttt{\smaller with}} \textit{*Exp}]) or values (\textit{pattern} and \textit{expr}), and if applicable, followed by parent constructs ([\textbf{\texttt{\smaller of}} *]). 
Expression constructs (rules~18-23) allow combining child elements with conjunctions or disjunctions (\textit{op}, rule 24) and parentheses.
The \textit{pattern} rule (rule~25) defines the structure for matching different parts of a pattern (rule~26) used for identifiers and keywords.
The \textit{expr} rule (rule~27) enables flexibly matching Java expressions. 
For example, a design rule might have the form: 
\begin{quoteRule}
\textbf{\texttt{\smaller class with name}} \textit{pattern} \textbf{\texttt{\smaller must have function with specifier}} \textit{pattern} \textbf{\texttt{\smaller and return value}} \textit{expr}
\end{quoteRule}
such as:
\begin{quoteRule}
\textbf{\texttt{\smaller class with name "...Cls" must have function with specifier "static" and return value "new ArrayList<String>()" }}
\end{quoteRule}
The language can be easily extended additional rules to the grammar and describing how each rule is mapped to an equivalent XPath query.

\subsubsection{Logical Connectives}
One challenge with representing a design rule through partial code snippets is that there is no natural way to express how partial code snippets should be combined. 
In some cases, the {\GUI} is able to resolve ambiguity by providing a default and enabling developers to change the default (e.g., by toggling a condition between a quantifier and constraint). However, in other cases it is clearer to express the meaning of the rule in a textual representation. 
One specific challenge is how to combine conditions: should including multiple conditions be interpreted as a conjunction or disjunction of conditions? By default, RulePad uses conjunction.
The {\TE} makes this choice explicit to the developer by inserting an 'and' terminal between conditions and enables them to edit the rule text to switch `and' to `or'. In addition, developers can use parentheses in the {\TE} to resolve association ambiguities.  

\subsubsection{Autocomplete}

Design rules written in the {\TE} must follow the grammar. To enable developers to more easily learn the grammar, developers may first use the {\GUI} to author design rules and then view the corresponding textual representation. In addition, the {\TE} offers autocomplete. After writing a token, autocomplete suggests potential next tokens that are valid given the grammar. This enables developers to see a list of valid completions they may write next.

The parser is unable to parse unfinished design rule text. Consequently, the parse tree is not available for autocomplete. In these cases, the {\TE} uses the information in the text and a set of rules to suggest valid next tokens. 
While providing suggestions for some cases is trivial (e.g.  `must' is always followed by `have'), providing suggestions for other cases requires more logic. For example, the list of available tokens after `must have' includes the AST children of the EoI, the first token in the text. However, if the EoI is a 2-part word (e.g. \texttt{\small declaration statement}), considering only the first token is not useful. In another example, `with' may be followed by possible AST children of the second to the last token.

If there is no autocomplete suggestion for the input text, the {\TE} checks the input text against the grammar, offering an error message for simple grammar violations. For example, if a developer writes a design rule with two instances of `must', an error message indicates that only one `must' is allowed, and the second `must' is decorated with a red wavy underline. 


\subsubsection{In-Context Documentation}

\begin{figure}
    \centering
    \includegraphics[width=0.45\textwidth]{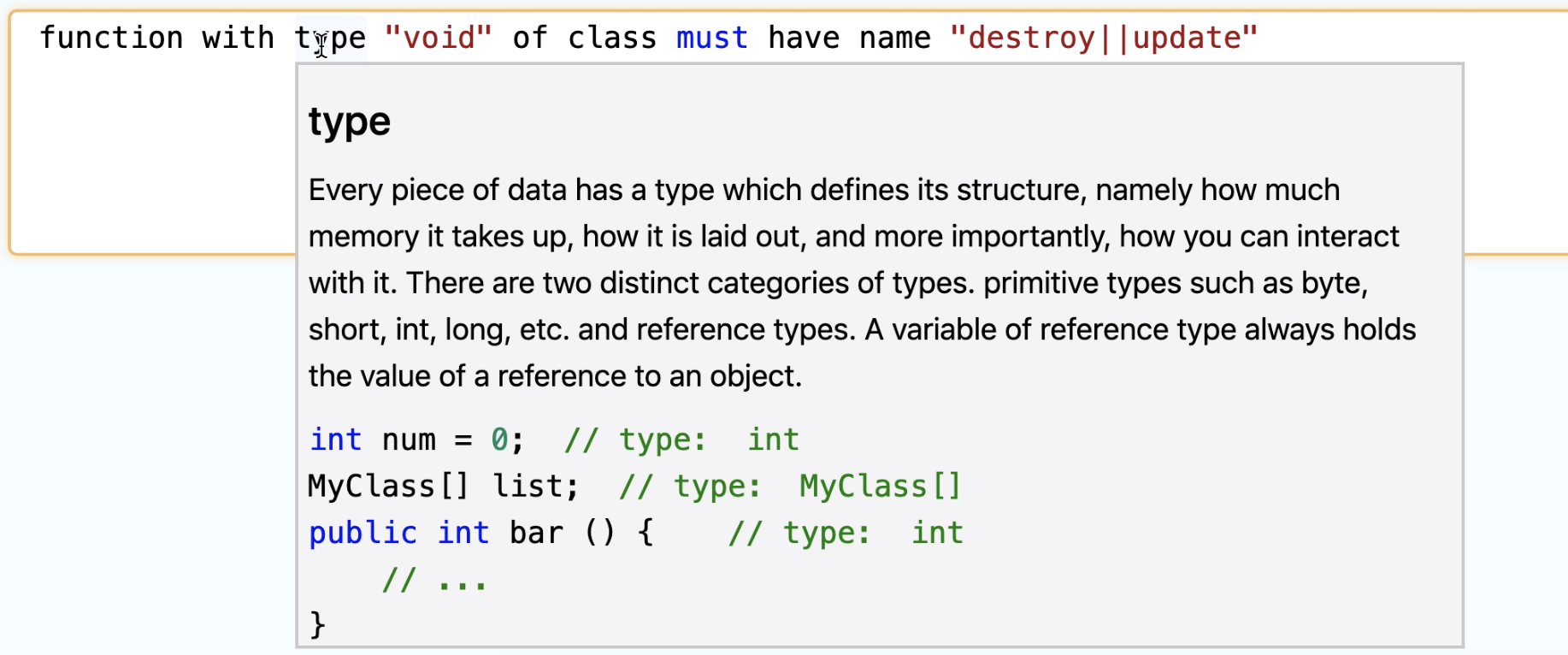}
    \caption{After moving the cursor over the word \textit{type}, in-context documentation describes its meaning and usage.}
    \label{fig:context_specific_documentation}
\end{figure}

When the developer moves their cursor over a word in the design rule, the developer is offered in-context documentation with an explanation of the word (Figure~\ref{fig:context_specific_documentation}). This enables learning the meaning of element names and keywords through descriptions and simple examples.  

\subsubsection{Linking the {\GUI} to the {\TE}}
Moving the cursor over an element name in the {\TE} highlights the corresponding element in the {\GUI} with a corresponding color. This enables the developer to easily edit rules in either representation and facilitates using both together to understand the current design rule.

\section{Evaluation}\label{sec:evaluation}

To evaluate our approach for authoring design rules, we conducted 
an experiment comparing the experiences of developers authoring design rules in RulePad and in PMD\footnote{The tool and the study materials are available at \href{https://github.com/devuxd/RulePad}{https://github.com/devuxd/RulePad}.}. The independent variable was the tool used by participants (RulePad or PMD), and the dependent variables included task time, errors, and reported willingness to use the tool in their own work.
We chose to compare RulePad against PMD,
a widely used rule checker.
Rule checkers, such as PMD, offer developers the ability to author custom rules expressing constraints to be checked against code. 
We selected PMD as PMD rules can be written using XPath, which RulePad uses internally to check rules, enabling a direct comparison.
In addition, no prior studies have examined the challenges developers may experience in authoring rules in PMD or other similar tools. 
PMD offers the Designer utility to write rules in XPath. Developers supply an example code snippet to which the rule can be applied, for which Designer then displays an AST. By referencing this generated AST, the developer may then write an XPath query expressing the desired rule (Table \ref{tab:XPath_example}).
The Designer utility then executes the XPath query against the supplied example code snippet, describing which parts, if any, of the supplied code snippet violate the rule. 

\subsection{Method}
To recruit participants, we invited students in relevant courses, posted flyers on bulletin boards,
and sent email invitations to Computer Science masters and Ph.D. students.
14 senior undergraduate and graduate students at our institution participated in the study. All had prior experience in Java. 
Participants had a median of 5 years of programming experience  (range 1.5 to 10) 
and 1.5 years of experience as a professional software developer (range 0 to 4).
Participants reported using a variety of programming tools, including text editors (Vim, Atom, Sublime Text, Evernote), version control systems, and issue trackers (Git). Control participants reported a range of levels of experience with ASTs and XPath. Three (C3, C1, C4) reported no prior knowledge, two (C6 and C2) reported some knowledge of ASTs but none of XPath, one (C7) reported limited knowledge in XPath but none in ASTs, and one (C5) reported some knowledge in XPath and reported having enough knowledge in ASTs. 
Each participant was compensated with a \$40 gift card for the 2-hour study session.

We used stratified random sampling to assign participants to one of two conditions, a control group (C1-C7) and an experimental group (E1-E7), balancing the levels of overall and professional experience reported in the pre-study survey.
Participants in both conditions had a similar distribution of programming experience (median 5 years for both groups) and professional experience (median 1 years for control and 1.5 for experimental). 

Participants in both groups were given an introduction about design rules and how design rules can be expressed using an IF/THEN structure. Then participants were asked to extract at least 3 design rules from the given source code. All were given 20 minutes to finish this task. 
Participants were then given a brief tutorial introducing RulePad or the PMD Designer (Version 6.12.0).
Experimental participants were given access to RulePad and control participants were given access to Designer and allowed to use the Internet.
Participants were given 80 minutes to complete the main task. As participants worked, we asked them to think aloud and captured a screen and audio recording. At the end of the study, we collected the final authored design rules and conducted a semi-structured interview. 
We asked participants about the features they found useful, their challenges, ideas for making the tool better fit their needs, and their willingness to use the tool in their everyday work.

\subsection{Tasks}

We adapted the source code used in the study from a simple Java project using the model-view-controller architecture\footnote{\href{https://github.com/derickfelix/BankApplication}{https://github.com/derickfelix/BankApplication}}. We chose the project as it is complex enough to contain several design rules yet simple enough that the logic of the code can be easily understood.

To familiarize participants with the source code and practice expressing rules in the IF/THEN structure, participants first completed a training task. Participants in both groups were asked to extract at least 3 design rules from the source code and write a textual representation of them in an IF/THEN structure by hand, without the use of any tool. All participants completed the training task (median 16 minutes). In the training task, all except E1 and C5 successfully expressed design rules in the IF/THEN structure.

In the main task, control and experimental participants were asked to use their tool to author 3 specified design rules, and one rule of their choice from the rules they extracted in the training task. The three specified design rules were given in increasing levels of complexity, evaluating the capability of each tool to support authoring rules of increasing complexity.
The first rule exercised basic features, the second rule exercised features for authoring complex rules with several constraints, while the third rule required using pattern matching expressions as well as possibly changing the element of interest
(Table~\ref{tab:study_2_rules}).

\subsection{Results}\label{sec:results}



In the main task, all 7 experimental participants were able to begin authoring at least one design rule (containing at least one query element), while 4 of the 7 control participants were unable to begin authoring any design rule.  
Participants in both conditions had up to 80 minutes to complete the main task. 
Experimental participants finished the task in a median of 60 minutes (mean 58 minutes, SD=15 minutes). 
All control participants were still working at the end of the allotted task time, except C2 who gave up after 62 minutes. Using a Shapiro-Wilk test, we were not able to verify that task time data is normal, and we thus used a Mann Whitney U test.
Participants with RulePad were significantly faster ($U=3$, $p<0.0074$, $r=0.76$). 

\renewcommand{\arraystretch}{.9}
\begin{table}[t]
\centering
\small
\caption{For each design rule (top), we identified a set of required query elements (numbered items in design rules and listed below) and scored each design rule on the correctness of each query element. The last row lists the mean score across all required query elements for all rules. 
}
\begin{tabular}{
p{0.27\textwidth}
>{\centering\arraybackslash}p{0.07\textwidth}!{\color{gray!70}\vrule width .5pt}
>{\centering\arraybackslash}p{0.07\textwidth}
}
& {Control} & {Experimental} \\  		
&	{Percentage Successful}	&	{Percentage Successful}	\\ 
\noalign{\global\arrayrulewidth=.7px}
\hline
\multicolumn{3}{p{.46\textwidth}}{\cellcolor{blue!10} \textbf{Rule I: } Model (1) classes must have (3) ``private'' (2) fields and (3)(4) getters.}								\\	
{	(1)	class	}	&	14\%	&	71\%	\\	
{	(2)	class field	}	&	14\%	&	71\%	\\	
{	(3)	private class field	}	&	0	&	71\%	\\	
{	(4)	class method	}	&	0	&	57\%	\\	
{	(5)	class method name	}	&	0	&	57\%	\\	
{	(6)	class method name ``...get''	}	&	0	&	57\%	\\	
{	No Extra Query Elements	}	&	14\%	&	57\%	\\	
\noalign{\global\arrayrulewidth=.5px}
\arrayrulecolor{gray!70} \hline
{	Overall Score	}	&	2\%	&	63\%	\\	
\noalign{\global\arrayrulewidth=.7px}
\arrayrulecolor{black} 
\hline
\multicolumn{3}{p{.46\textwidth}}{\cellcolor{blue!10} \textbf{Rule II: } Repository (1) classes except (2)(3) ``BaseRepository''  must (4) extend (5) ``BaseRepository'' and (6) implement an interface and they should have a (7)(8) ``...Mapper'' (9) function.}								\\	
{	(1)	class	}	&	14\%	&	100\%	\\	
{	(2)	class name	}	&	14\%	&	14\%	\\	
{	(3)	class name except ``BaseRepository''	}	&	0	&	14\%	\\	
{	(4)	class implements	}	&	0	&	86\%	\\	
{	(5)	class extends	}	&	0	&	86\%	\\	
{	(6)	class extends ``BaseRepository''	}	&	0	&	71\%	\\	
{	(7)	class method name	}	&	0	&	100\%	\\	
{	(8)	class method name ``...Mapper''	}	&	0	&	71\%	\\	
{	(9)	class method	}	&	0	&	100\%	\\	
{	No Extra Query Elements	}	&	0	&	57\%	\\	
\noalign{\global\arrayrulewidth=.5px}
\arrayrulecolor{gray!70} \hline
{	Overall Score	}	&	3\%	&	70\%	\\	
\noalign{\global\arrayrulewidth=.7px}
\arrayrulecolor{black} 
\hline

\multicolumn{3}{p{.46\textwidth}}{\cellcolor{blue!10} \textbf{Rule III: } (1) Name of (2)(3) ``void'' (4) functions in Controller classes are either (5) ``store'', ``update'', ``deposit'', ``withdraw'', or ``destroy''.}								\\	
{	(1)	class method name	}	&	0	&	14\%	\\	
{	(2)	class method return type	}	&	0	&	0	\\	
{	(3)	class method return type ``void''	}	&	0	&	0	\\	
{	(4)	class method	}	&	14\%	&	14\%	\\	
{	(5)	class method name ``store, update, deposit, withdraw or destroy''	}	&	0	&	0	\\	
{		No Extra Query Elements	}	&	0	&	29\%	\\	
\noalign{\global\arrayrulewidth=.5px}
\arrayrulecolor{gray!70} \hline
{	Overall Score	}	&	2\%	&	10\%	\\
\noalign{\global\arrayrulewidth=.7px}
\arrayrulecolor{black} 
\hline \hline
Mean score of required query elements for Rules I, II, and III & 4\% & 52\% \\
\hline
\end{tabular}
\label{tab:study_2_rules}
\end{table}

\subsubsection{Completeness and Correctness of Design Rules} 
To assess the completeness and correctness of the design rules participants wrote, we scored each design rule. To compare design rules written using the PMD designer to those written using RulePad, we first took each design rule authored in RulePad and extracted the resulting XPath query generated by RulePad.  
We then scored each response, identifying each required query element necessary in the AST paths defined by the XPath query. Table~\ref{tab:study_2_rules} lists the required query elements for each design rule. 
For each query element, if the design rule included the necessary AST nodes in the correct format and order, the query element was scored as correct. In addition, we marked design rules which included irrelevant query elements. 

Across all required query elements in Rules I, II, and III, experimental participants on average were successful in correctly authoring 52\% of the elements. In contrast, control participants succeeded in writing 4\% of the elements (Table~\ref{tab:study_2_rules}).
We used a Shapiro-Wilk test to test the normality of the design rule scores. We were not able to verify the normality of our data, and we thus again used a Mann Whitney U test.
Experimental participants authored significantly more correct query elements for Rules I ($U=0$, $p<0.0022$, $r=0.86$)
and II ($U=2$, $p<0.0004$, $r=0.84$). 
Experimental participants wrote more correct query elements for Rule III, but the difference was not significant ($U=11.5$, $p>0.3371$, $r=0.36$). 
Rule III was more complex, and the six rules experimental participants authored for it did not include all of the required elements.

As Rule IV was selected by each individual participant and varied between participants, we identified required query elements necessary for each separate design rule. On average, design rules authored by participants with RulePad received a score of 72\%. Design rules authored by participants in PMD received a score of 0. No control participant succeeded in authoring any part of Rule IV.

\subsubsection{Perceived Value of Tools}
In post-task interviews, we asked participants about their willingness to use the tool they used within the study in their everyday work. On a 7-point Likert scale (1 being the least preferred), experimental participants reported being significantly more willing to use RulePad (mean 5.2, SD=1.5) than control participants to use the PMD Designer (mean 3.7, SD=1.1)(one-tailed equal t-test, $p<0.023$). Control participants reported that the main reasons they were willing to use PMD and PMD Designer were its perceived popularity and ability to find defects.

\subsubsection{Challenges using the PMD Designer}
Only 3 of the 7 control participants were able to author any design rule at all, authoring a total of 4 design rules.
Participants using the PMD Designer faced several significant challenges using the tool and its documentation which hindered their progress.

\textbf{Unclear Expected Notation: } 
All control participants experienced challenges determining the type of input (e.g. text, code snippets, class files, etc.) the tool expected them to enter to author a design rule.
The Designer includes a main panel for writing example Java code. However, this was not discoverable by many participants. 
C3 expected the Designer to be an IDE and tried to write design rules in Java rather than XPath. C7 assumed the tool expected them to enter XML template text, as used in the official website. 
The Designer user interface consists of three read-only and three editable input boxes. This distinction was unclear. Only two participants (C1 and C4) successfully discovered the component for writing and wrote XPath queries. After using Designer to compute the AST of a sample code snippet, C6 instead directly authored the XPath query in the XML file executed by PMD.
C3 tried to import the project into the tool.

\textbf{Inadequate Documentation: }
Participants experienced challenges using the steps described in the official documentation due to a lack of detail.
Participants reported they were confused about steps and required actions (C4). 
For example, after the AST is displayed, some participants assumed the rule had already been generated and were confused to see the fields of imported XML text not filled in.
At the same time, some (C7, C6) reported that the official tutorial was too long and overwhelming for new users.
This tutorial includes information about writing XPath queries. This information misled one participant, who spent considerable time looking for an answer to a specific question
about XPath 
on the website (C1).
Participants wished to instead see documentation which included a `Hello World' example (C7), integrated tutorial (C2), description of what developers needed to do (C4), and a visual tour for new users (C6).




\textbf{Poor Error Handling: }
In the PMD Designer, compilation errors for source code and XPath queries are reported through a small icon at the top of the component. In the post-task interview, C4 reported that these were not visible.
XPath query matches as well as error messages are reported in the component adjacent to the XPath query field. 
Compilation error messages did not provide information about how to fix errors. 
C1 suggested the need for better error handling, in particular localizing the line where the error occurs. 

\subsubsection{Experiences and Challenges using RulePad}


The in-tool tutorial, dynamic guide, and walk-through tutorial helped familiarize participants with the steps for authoring a design rule.
All except E4 used the dynamic guide to determine the requested actions in each step. 
In contrast, only three (E1, E4, E5) read the walk-through tutorial given when they first opened RulePad. 
All participants used the {\TE} to verify design rules. E7 and E6 initially authored a design rule in the {\TE}. But after receiving an error, they continued instead with the {\GUI}. After completing the steps in the {\GUI}, they used the {\TE} to understand the grammar for expressing rules. None of the participants tried to author rules from scratch using the {\TE}. 

In the post-task interview, participants reported several features which helped them to author design rules, including the unambiguous nature of the design rule template (E2, E4, E6, E7), the ability to specify characteristics explicitly (E4), error handling (E6), the pop-up tutorials 
in the {\GUI} 
(E2), automatically ensuring the structure of the rule in the code (E5), and presenting the design rules in a readable form (E1).

The study revealed several usability issues. Some participants misinterpreted the meaning of some of the AST elements in the GUI (E1, E6). 5 participants added extra query elements to at least one design rule. 3 of the 7 experimental participants were unable to successfully use pattern matching. 
To help users reuse data, RulePad initially populates the template with the data of the previously authored rule and offers a `Clear Form' button at the bottom of the form. This confused participants, and the `Clear Form' button was not discoverable by some participants. 
Some participants were confused by the first step of the dynamic guide, as it prompted them to specify conditions on quantifiers. However, the first design rule did not require a condition on the quantifier.  
The paths of the files and folders to which the design rule applies follows a standard format with slashes. As there is no validation for these paths, E5 and E6 entered incorrect paths, and E2 and E3 used an incorrect format for specifying paths without receiving any error. As a result, no matches or violations were listed, and participants incorrectly assumed that the tool did not check rules against the code or offer immediate feedback.

\section{Limitations and Threats to Validity}

As in all studies, our study has several important limitations and threats to validity. 
Participants were artificially limited in learning about the tools exclusively by working with the tools and their documentation. In practice, developers may have access to other resources or to experienced teammates. However, developers often rely first on building their own understanding, as they may be expected to make an effort to understand before engaging with teammates~\cite{latoza2006maintaining}.

Unlike developers working in their own codebase, participants were unfamiliar with the code they worked with. In practice this situation may occur when developers are new to a team or find themselves working in part of a large codebase with which they are previously unfamiliar.
Much of the task involved working with a specific persistence framework, with which developers were unfamiliar. Developers with more experience in a codebase or a framework might benefit less, as they might have internalized more of the design rules. 
At the same time, as participants wrote design rules and viewed the matching code, participants were unable to use feedback to refine their rules.
None knew what the correct behavior of the rule should be, and were thus unable to use the list of matches and violations to judge the correctness of their rule. In practice, we would expect developers authoring a design rule to have a clear expectation of which code snippets should and should not be violations and use this feedback to identify issues with the rule to be resolved.
This may have made writing correct rules artificially difficult.


Our selection of participants may threaten the generalization of the results. To mitigate this threat, we recruited participants with a range of experience levels, from one to ten years of industrial experience. In the control group, participants with prior experience working with AST and XPath queries performed slightly better than other participants. Participants with more prior experience with program analysis, or PMD in particular, would likely have performed better.

\section{Related Work}\label{sec:related_work}
A variety of tools have been designed which use a catalog of rules to check for potential defects in code.
PMD~\cite{copeland2005pmd} and FindBugs~\cite{hovemeyer2004findbugs} are widely used tools used for finding statement-level defects. CheckStyle helps developers identify coding style defects~\cite{burn2005checkstyle}. 
ActiveDocumentation checks code for conformance against design rules and reports matches and violations in code~\cite{Mehrpour2019ActiveDoc}. 
While most enable developers to write custom rules for their own project, they require developers to interact with specialized query notations or a program analysis framework to do so.


Other tools support analyzing the architectural structure of code~\cite{passos2010static} through various techniques, including visualizing or querying the architecture of the code. Visualization techniques such as dependency-structure matrices (e.g., Lattix DM~\cite{Sangal2005lattix}) and reflexion models (e.g., SAVE~\cite{knodel2007comparison}) display the software architecture through matrices and graphs and allow developers to compare them against the intended architecture. Source code query languages (e.g., Semmle.QL~\cite{deMoor2008}) allow developers to query the intended architecture of the code and report textual lists, graphs, and diagrams.

Source code query languages (SCQL) and tools enable developers to query source code 
to check conformance to potential design rules or find patterns in code.
Some offer query languages which are close to natural language~\cite{srcQL-BartMan,BBQ}. 
srcQL is a code query language built upon the XML representation of the AST and XPath and is modeled on SQL~\cite{srcQL-BartMan}.
Developers are able to query code using a semi-structured natural language in Browse-By-Query (BBQ)~\cite{BBQ}.
SOUL~\cite{DeRoover:2011:STS} and Rscript~\cite{klint2003understanding} provide functional languages.
Unlike RulePad, many source code query tools require knowledge of the internal AST to write a query and full compilation of the code to execute it ~\cite{deMoor2008,DeRoover:2011:STS,klint2003understanding}. More broadly, SCQL tools are specialized for querying code rather than authoring rules.

Studies have investigated the barriers developers face when working with program analysis tools as well as offered approaches to mitigate those issues.
Developers face a number of barriers that can lead to misuse of the tools and discourage them from using them~\cite{ChristakisDevelopersWant,Johnson:2013:WDS}. 
These include a high number of false positives, poor organization of warnings, lack of support for collaborative environments, disconnection from developers' workflow, and difficulties in interpreting the results.
Developers report that the integration of tools into their development environment, organized warnings, and immediate feedback are helpful and motivate them to write better code during code reviews~\cite{Tymchuk:2018,Johnson:2013:WDS}.
Other approaches include customizing program analysis tools at the project level, which can be more successful than incorporating customization at the user level~\cite{Sadowski:2015:TBP}. 


Natural language interfaces enable users to directly express queries without learning system commands or interfaces~~\cite{Srinivasan:2017:NLI:3308567.3308579}. One early focus was in introducing natural language interfaces to generate database queries (e.g.,~\cite{androutsopoulos,Popescu2003theoryOfNLI}). 
PRECISE~\cite{popescu2004modern} translates natural language inquiries into corresponding SQL queries by matching tokens discovered in the user query with the database schema.
NaLIX takes an English sentence as a query and converts it to an XML query executable on XML databases~\cite{li2005nalix}. 
Other work introduced natural language interfaces to support data exploration in information visualization applications~\cite{cox2001multi}.
One example is \textit{Articulate}, which leverages natural language processing to translate users' crude verbal descriptions into actionable and valid visualization expressions, without requiring a user to learn a complex user interface or query language~\cite{sun2010articulate}.

A variety of interactive systems have explored approaches for enabling users to interactively refine query text to add necessary details by continuously offering feedback as the user works.
Eviza supports users in creating an interactive session with data using natural language and a probabilistic grammar~\cite{setlur2016eviza}.
It enhances natural language queries by first guessing the user's intent and then offering ambiguity widgets to change the guess.
DataTone offers a mixed-initiative approach for handling ambiguity in natural language interfaces in the context of data visualization~\cite{gao2015datatone}.
NLyze offers a robust natural language interface for creating formulas in spreadsheets~\cite{gulwani2014nlyze}.
Orko utilizes both natural language and direct manipulation of input for network visualization and uses graphical widgets such as dropdowns to resolve ambiguities~\cite{srinivasan2018orko}. FREyA reduces the need for customizing domain-specific languages for each domain 
by engaging the user in clarification dialogs~\cite{Damljanovic2010}.

\section{Discussion}\label{sec:discussion}


In this paper, we introduced two techniques for authoring checkable design rules: snippet-based authoring and semi-natural-language authoring. 
These enable developers to work with design rules in more natural and easy to learn representations.
Compared to authoring rules in PMD, 
we found that developers were able to successfully author 13 times more query elements in significantly less time and wrote query elements which were 48\% more correct.
Developers working with RulePad reported significantly more willingness to use RulePad in their everyday work than developers working with PMD.
Developers using RulePad reported benefiting from the unambiguous nature of the design rule template and the ability to specify characteristics explicitly and interactively. Developers used the synchronized {\TE} to review and better understand the rules they constructed. 

Semi-natural language interfaces can sometimes lead to confusion, as it may be hard to learn the limits of what can be expressed in the language as well as constraints imposed by its syntax and grammar on how it should be expressed.
RulePad offers several mechanisms to reduce these issues.
First, developers may begin writing rules using the {\GUI}. Second, autocomplete offers a way to discover what can be expressed. Third, the bidirectional connection of the editors enables developers to use rules they have already written using the {\GUI} to understand how to express them using the {\TE}, and vice versa. Finally, immediate feedback allows developers to see if the rule behaves as intended by examining the matches and violations generated by the rule.


Our study offers evidence for the challenges developers face in customizing a new program analysis tool for their work context. 
While we expected the difficulty of authoring XPath queries to be the primary challenge, we found that participants were often confused long before they reached that point. Usability issues working with the tool were often a significant challenge for participants. While diligent and dedicated developers may ultimately succeed, these issues offer a significant barrier dissuading developers from adapting it to their project. 

 


There are a number of ways in which our semi-natural language might be made even closer to natural language. One approach might be to enable developers to define their own vocabulary, giving developers more flexibility in how rules are expressed. This might be achieved through macros. Developers might take a series of tokens in a design rule (e.g, a declaration statement with visibility "private") and offer a short form for this text (e.g., "private fields"). Design rules might then be written and displayed to the user with a mixture of grammar and short form text.  Short form text might be saved and shared across design rules, creating a dictionary of alternative terms for expressing common ideas. Autocomplete might assist developers in discovering these short forms. 

%

While our system focuses on authoring design rules which may be expressed as a pattern in an abstract syntax tree, our approach might be applied to authoring design rules which offer constraints over other representations of programs and their execution. The key insight of snippet-based authoring is that, instead of learning a complex notation, developers can instead give examples of what they wish to match and then interactively refine the rules generated from this to resolve any ambiguity. This method of interactive example refinement might be used to construct other forms of specification. 
For example, ordering relationships might be expressed by listing a sequence of method invocations which must occur.  
To reduce the ambiguity in interpreting a sequence of method invocations, developers might interactively indicate that some are optional or may occur multiple times. 
As usability is often a key barrier in adopting specification languages, we believe that interactive example refinement is an approach which may be more widely applicable.

Working with design rules might also be integrated into code transformation systems. If, for example, a developer edited a code snippet so that it no longer satisfies a design rule, the developer might receive feedback that their code has violated a rule. Developers might then be given a choice: fix their code to match the rule or update the rule itself. Each of these actions might be automated in whole or in part through a code transformation system. Updating the rule might initiate a refactoring, where the updated rule is used to transform every instance to which it applies.


\section*{Acknowledgements}
We thank our study participants for their time. This work was supported in part by the National Science Foundation under grant NSF CCF-1703734.

\balance

\bibliographystyle{ACM-Reference-Format}
\bibliography{bibs}

\end{document}